\documentclass[a4paper,11pt]{article}

\usepackage{pos}

\title{Universal and non-universal finite-volume effects in the vicinity of chiral phase transition in (2+1)-flavor QCD}
\ShortTitle{Universal and non-universal features of (2+1)-flavor chiral phase transition}

\author*[a]{Sabarnya Mitra}
\author[a]{Jishnu Goswami}
\author[a]{Frithjof Karsch}
\affiliation[a]{Fak{\"u}ltat F{\"u}r Physik, Universit{\"a}t Bielefeld, D-33615 Bielefeld, Germany}

\emailAdd{smitra@physik.uni-bielefeld.de}
\emailAdd{jishnu@physik.uni-bielefeld.de}
\emailAdd{karsch@physik.uni-bielefeld.de}

\def\lsim{\raise0.3ex\hbox{$<$\kern-0.75em\raise-1.1ex\hbox{$\sim$}}}
\def\gsim{\raise0.3ex\hbox{$>$\kern-0.75em\raise-1.1ex\hbox{$\sim$}}}

\newcommand{\LA}{\left \langle}
\newcommand{\RA}{\right \rangle}

\abstract
{
In this proceeding, we discuss the finite-size scaling analysis of the order parameter related to the chiral phase transition in QCD with two massless quarks. We use data obtained in lattice QCD calculations performed with highly improved staggered quarks (HISQ) for a range of light quark masses, $1/240 \leq m_\ell/m_s \leq 1/27$ for different spatial volumes ($N_\sigma$) on Euclidean lattices with temporal extent $N_\tau=8$, satisfying $3\,N_\tau \leq N_\sigma \leq 10\,N_\tau$\,.
We observe that infinite volume extrapolated data
for the order parameter agree reasonably well with the expected $O(2)$ scaling behavior even for
physical ratios of the light-to-strange quark mass
ratio. We quantify deviations from asymptotic 
scaling and perform a detailed analysis of the 
influence of
finite-size effects in terms of temperature and quark masses at a fixed lattice cutoff. This is crucial for improving the reliability of the infinite-volume extrapolated estimate of the chiral order parameter and for a more precise determination of chiral phase transition temperature from direct Lattice QCD simulations.
}

\FullConference{The 42nd International Symposium on Lattice Field Theory (LATTICE2025) \\
2-8 November 2025 \\
Tata Institute of Fundamental Research, Mumbai, India
\\
}

\begin{document}
\maketitle

\section{Introduction}

Despite having commendable lattice Quantum Chromodynamics (QCD) evidences for the restoration of spontaneously broken chiral symmetry, $SU(2)_L\times SU(2)_R$, in the limit of vanishing light quark masses through a second order phase transition\footnote{For a recent overview and further references, see
\cite{Schmidt:2025ppy}.} 
%along with somewhat known broken nature of axial $U(1)_A$ symmetry at the chiral phase transition temperature $T_c$
along with the explicitly broken nature of the $U(1)_A$ symmetry at the QCD transition temperature \cite{HotQCD:2012vvd,Ding:2020xlj}, 
the role of the axial anomaly in the vicinity of the 
chiral phase transition, in particular the strength of its breaking and its impact on physical QCD thermodynamic observables continues to
remains to be an open question. It is therefore quite crucial to provide further concrete support for the universal critical behavior of chiral phase transition, which in the limit of two vanishing light quark masses does
resemble the universality class of the $3$-dimensional, $O(4)$
spin model \cite{Pisarski:1983ms}, and which is also supposed to provide stringent constraints on the location of the elusive QCD critical point in the QCD phase diagram \cite{Ding:2024sux,Goswami:2025wtj}.
Although several studies, comparing lattice QCD calculations with expected universal critical behavior close to the chiral limit, are already existent in the related state-of-the-art literature, a majority of these numerical studies
base their point for observation of universal critical behavior on the very presumption that the transition belongs to the $3$-d, $O(4)$ universality class and on subsequently
constructing the corresponding universal scaling ansatz 
to parameterize deviations away from criticality
in terms of additional non-singular terms contributing to the relevant observables. Thus, a direct lattice QCD determination of the critical parameters influencing the QCD chiral phase transition obviously is desirable. 

In this work, we systematically approach towards a direct lattice QCD based calculation of the critical universal parameters related to the QCD chiral phase transition. We remark that we will not yet aim at presenting continuum
extrapolated results in this work. 
We rather focus on a detailed analysis of finite volume and non-vanishing quark mass effects on chiral observables that eventually will allow a direct determination of universal amplitudes and critical exponents in QCD, without relying on any form of presumption regarding the underlying universality class. For recent developments and progress already made, kindly refer to our previous works \cite{Mitra:2024mke,Mitra:2025aeu,Mitra:2025hsk}.
Working within the domain of zero density here, we will analyze an improved order parameter for quantifying the extent of chiral symmetry restoration in (2+1)-flavor, lattice regularized QCD for a non-zero value of the lattice cut-off or spacing. 

The proceedings is organised as follows: 
In the next Section \ref{sec:theory}, we briefly introduce the basic quantities of relevance for this work along with the core theory concerning the finite-volume analysis of QCD chiral phase transition. In Section \ref{sec:Results}, we briefly describe the setup and present the main results of our current study. We summarise the present state of the work in Section \ref{sec:Summary}.

\section{Finite-size scaling analysis of (2+1)-flavor QCD chiral phase transition with HISQ fermions}
\label{sec:theory}

In this section we present a brief summary of the basic equations \cite{Karsch:2023pga,Engels:2014bra,Engels:2001bq} relevant for finite-size scaling analysis of chiral phase transitions. It is well-known that staggered fermion discretization preserves only part of the full $SU(2)_L\times SU(2)_R$ chiral symmetry on the lattice. For (2+1)-flavor QCD at finite lattice spacing, the remnant chiral symmetry group is associated with the $3$-d, $O(2)$ universality class, which governs the universal critical behavior close to the chiral phase transition. In the limit of vanishing light quark masses, the chiral condensate $M_\ell$ ($\ell=u,d$), and the chiral susceptibility $\chi_\ell$ are the order parameter and the corresponding susceptibility of the chiral phase transition \cite{Bazavov:2011nk}\,, defined as : 
\begin{equation}
    M_\ell=\frac{m_s}{f_K^4}\,\LA \bar{\psi}\psi \RA_\ell^{n_f=2}\;\;,\;\; \chi_\ell = m_s\,\frac{\partial M_\ell}{\partial m_\ell} = \frac{m_s^2}{f_K^4} \;\chi_{\ell}^{n_f=2} \; ,
    \label{eq:RGI}
\end{equation}
with $m_s$ denoting the strange quark mass, which is 
used to define a multiplicatively renormalized chiral
condensate, and appropriate factors of the
kaon decay constant $f_K$ to arrive at a dimensionless order parameter and its susceptibility.
To remove the ultraviolet (UV) divergence of the bare chiral condensate, we define a renormalization group invariant (RGI) improved order parameter in a temperature-independent normalisation,  
\begin{align}
M(T,H,L) = M_\ell - H \chi_\ell\;,
\label{eq:M}
\end{align}
which is a function of temperature $T$, light-to-strange quark mass ratio $H=m_\ell/m_s$ and the aspect ratio $L=N_\sigma/N_\tau$, where $N_\sigma$ and $N_\tau$ denote the spatial and temporal extent of the lattice, respectively. 
Chiral observables are obtained by taking derivatives with respect to $m_\ell$ of the (2+1)-flavor QCD partition function $Z$ given as :
\begin{eqnarray}
	Z(N_\tau,N_\sigma,m_\ell,m_s) = \int DU \; e^{-S_g[U]}\;\big[\det D_\ell(U,m_\ell)\big]^{1/2}
	\times\big[\det D_s(U,m_s)\big]^{1/4}  \; ,
	\label{eq:Z}
\end{eqnarray}
where $D_\ell$ and $D_s$ denote the light and strange quark staggered Dirac operator, respectively. 
We use the 2-flavor normalisation for the light quark chiral condensate and its susceptibility,
\vspace{.2cm}
\begin{equation}
\LA \bar{\psi}\psi \RA_\ell^{n_f=2} 
= \frac{T}{V}
\frac{\partial \ln Z}{\partial m_\ell} = 
\frac{1}{2} \frac{1}{N_{\sigma}^3 N_{\tau}} 
\big \langle \,{\rm Tr} \, D_\ell^{-1} \,\big \rangle \;\;\;,\;\;\;\text{and}\;\;\;\;\; 
\chi_{\ell}^{n_f=2} = \frac{\partial}{\partial m_\ell} \,\LA \bar{\psi}\psi \RA_\ell^{n_f=2}   
\label{eq:chiral}
\end{equation}
In the vicinity of the critical point, the order parameter $M_\ell$ and
the susceptibility $\chi_\ell$  are predominantly controlled by the so-called singular parts, which can be represented in terms of scaling functions 
$f_G$ and $f_\chi$, respectively. They are functions of scaling variables $z,z_L$, 
\begin{eqnarray}
	M_\ell(T,H,L) &=& h^{1/\delta} f_G(z,z_L) \; , \label{eq:M-sing}\\
	\chi_\ell(T,H,L) &=& h_0^{-1} h^{1/\delta-1} f_\chi(z,z_L)\; ,
	\label{eq:chi-sing}
\end{eqnarray}
where,
\begin{equation}
	f_{\chi}(z,z_L) = \frac{1}{\delta}\, f_{G}(z,z_L)
- \frac{z}{\beta\delta} \frac{\partial f_{G} (z,z_L)}{\partial z}
	-\frac{\nu z_L}{\beta\delta} \frac{\partial f_{G} (z,z_L)}{\partial z_L}~,
\label{eq:fchi}
\end{equation}
and $h=H/h_0$ with $h_0$ being a non-universal constant 
that sets the scale for the symmetry breaking parameter $H$.
Similar non-universal parameters $t_0, \ell_0$ appear as respective scales for the reduced temperature $\tau=(T-T_c)/T_c$ and the dimensionless volume\footnote{We denote by $L$, the spatial extent of a three dimensional lattice expressed in units of the inverse temperature, $L=V^{1/3}\,T$. In the context of lattice QCD, this is identical to the aspect ratio $L\equiv N_\sigma/N_\tau$.} $L$.
The scaling variables $z$ and  $z_L$ are defined as
\begin{eqnarray}
	z = z_0 \,z_b \;\; , \;\; z_L = z_{L,0}\,z_{L,b} 
	\; ,
\label{eq:zzL}
\end{eqnarray}
with $z_0=h_0^{1/\beta\delta}/\,t_0$\;,\; $z_{L,0}=h_0^{\nu/\beta\delta}\,\ell_0$\;, and the bare scaling variables $z_b\,,\,z_{l,b}$, obtained directly from the `bare' parameters 
entering any numerical computation are :
\begin{eqnarray}
	z_b= \frac{\tau}{ H^{1/\beta\delta}} \;\; ,\;\;
	z_{L,b}=\frac{1}{L H^{\nu/\beta\delta}}\; .
	\label{eq:zzL0}
\end{eqnarray}
The critical exponents $\beta$, $\delta$ are well-known for the  
$3$-d, $O(2)$ universality class \cite{Pelissetto:2000ek,Hasenbusch:2019jkj}, which we use,  
\begin{equation}
	\beta= 0.34864(7) \;\; ,\;\; \delta=4.7798(5) \;.
	\label{critical}
\end{equation}
Furthermore, the critical exponent $\nu$ entering the definition of $z_L$ is related to these two critical
exponents through a hyper-scaling relation,
\begin{equation}
	\frac{\nu}{\beta\delta} = \frac{1}{3}\;\left(1+\frac{1}{\delta}\right) \;.
	\label{eq:nu}
\end{equation}
Universal scaling functions for the order parameter and the order 
parameter susceptibility have been analyzed in quite some
detail for the $3$-d, $O(2)$ universality class (see e.g. \cite{Engels:2001bq,Springer:2015kxa}), although an explicit 
parametrization for finite volume scaling functions in this universality class has been given in terms of a Taylor series in $z$ and $z_L$, only recently 
\cite{Karsch:2023pga} : 
\begin{eqnarray}
%\hspace{-0.5cm}
f_{G}(z,z_L) &=& f_{G}(z) + 
\sum_{n=0}^{5}\sum_{m=3}^{8} a_{nm} \,z^n\, z^m_L
\; , 
\label{eq:fgchiV}
\end{eqnarray}
in the range  $z\in [-1:2]$ and $z_L\in [0.4:1.0]$
with coefficients $a_{nm}$ given in \cite{Karsch:2023pga}.  
In the infinite volume limit, $z_L=0$, the volume-dependent term of the 
scaling functions obviously vanishes and we obtain the standard one-parameter form of the infinite volume scaling function $ f_\chi(z)$.

\begin{figure*}[b]
\includegraphics[width=0.45\textwidth]{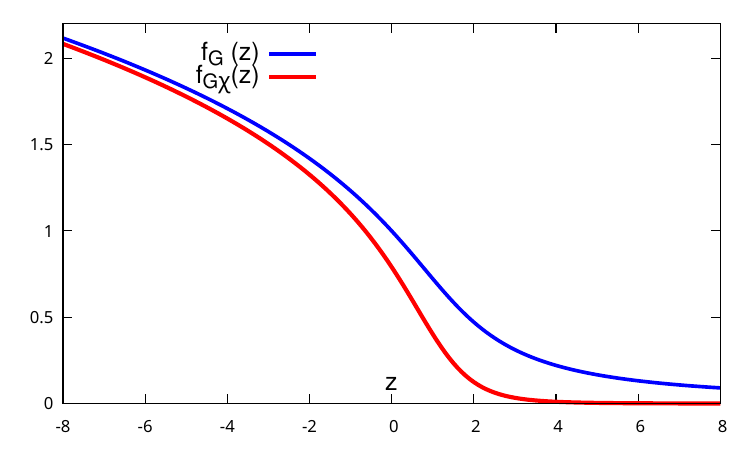}
\includegraphics[width=0.45\textwidth]{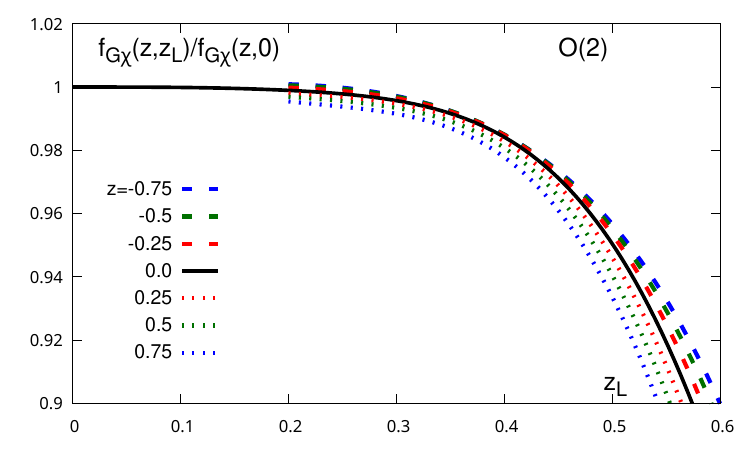}
\caption{{\it Left:} The $3$-d, $O(2)$ infinite volume scaling functions
$f_G(z)$ and $f_{G\chi}(z)$ versus $z$.
{\it Right:} The ratio of the $3$-d, $O(2)$ finite volume scaling function
$f_{G\chi}(z,z_L)$ and the infinite volume result,
$f_{G\chi}(z,0)\equiv f_{G\chi}(z)$,  versus $z_L$. }
\label{fig:fGfchi}
\end{figure*}

Using Eqs.~\eqref{eq:M-sing} and \eqref{eq:chi-sing} we find for the 
improved order parameter introduced in Eqn.~\eqref{eq:M} can be written as:
\begin{equation}
	M = h^{1/\delta} f_{G\chi}(z,z_L) \;\;,\;\text{with}\;\; f_{G\chi}(z,z_L)=f_G(z,z_L) - f_\chi(z,z_L) \; .
	\label{eq:MfG}
\end{equation}
In Fig.~\ref{fig:fGfchi}~(left) we show the
infinite volume order parameter scaling function
$f_{G\chi}(z)$ for the improved order $M$ and compare
it to the scaling function $f_G(z)$ for the 
conventional order parameter. Its finite-size 
dependence is shown in Fig.~\ref{fig:fGfchi}~(right)
for a few values of $z$ that are relevant for the
study of universal critical behavior close to $T_c$
and for the range of symmetry breaking parameters $H$
used in our current study.

For $z=z_L=0$, \textit{i.e.} at the critical point, the scaling function $f_{G\chi}(z)$ obeys the normalization $f_{G\chi}(0)=1-1/\delta$. The scaled order parameter $M/H^{1/\delta}$ at this critical point thus
is proportional to the non-universal constant $h_0^{-1/\delta}$,
\begin{equation}
	M(T_c,H)/H^{1/\delta} = h_0^{-1/\delta} (1-1/\delta) \; .
	\label{eq:Mh0}
\end{equation}
Also note that $M(T,H)/H^{1/\delta}$ will diverge
in the chiral limit for $T<T_c$ and vanishes for
$T>T_c$. This feature, arising from the asymptotic expansion of scaling function $f_{G\chi}$, actually is quite insensitive 
to a particular choice of the exponent $\delta$.
For $z=0$ \;\textit{i.e.} at $T=T_c$, the parametrization of finite volume corrections take on a simple
form in terms of a polynomial in $z_L$ only,
\begin{eqnarray}
    f_G(0,z_L)&=& 1+\sum_{m=4}^8 a_{0m} z_L^m
        \; , 
\end{eqnarray}
with  coefficients $a_{0m}$ given in Ref.~\cite{Karsch:2023pga}.

\section{Results}
\label{sec:Results}

\begin{figure}[b]
\includegraphics[width=0.45\textwidth]{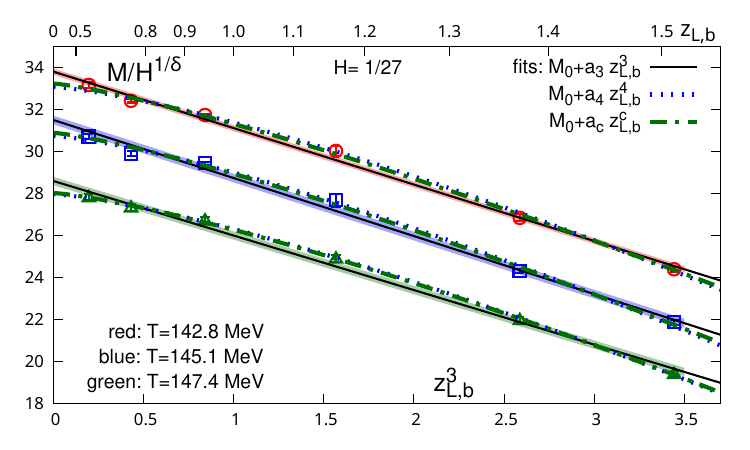}
\includegraphics[width=0.45\textwidth]{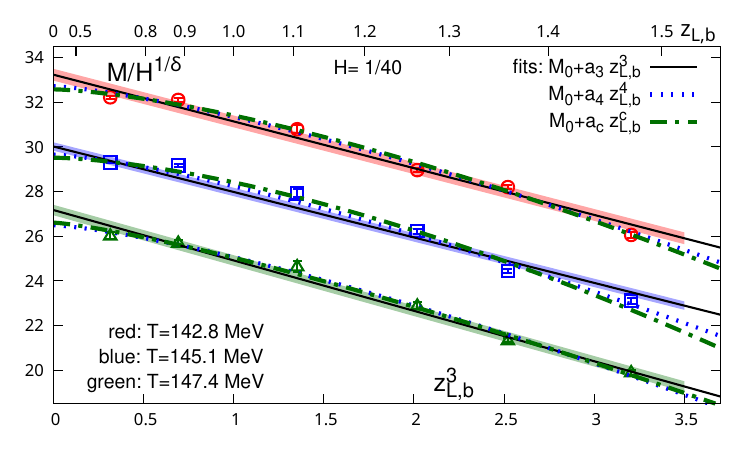}
\begin{center}
\includegraphics[width=0.45\textwidth]{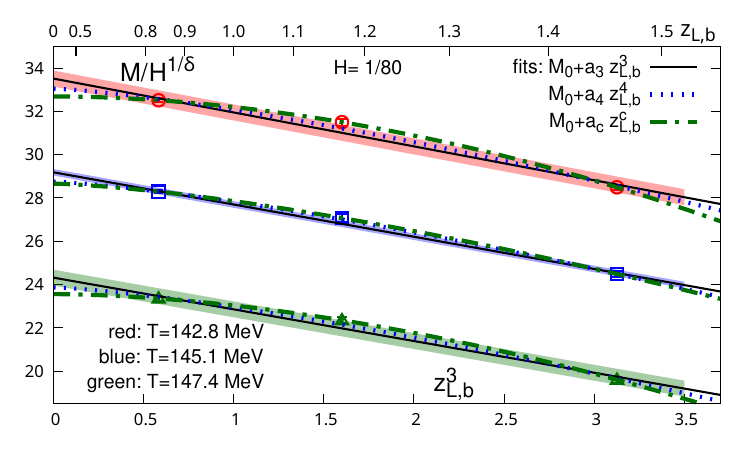}
\end{center}
%\hspace{1.5cm}
%\includegraphics[width=0.45\textwidth]{figures/Vcorr.pdf}
\caption{The scaled order parameter, $M/H^{1/\delta}$, for 
three values of the light-to-strange quark mass ratio, $H=1/27, 1/40$ and $1/80$, versus
the bare finite volume scaling variable $z_{L,b}$. 
}
\label{fig:Mscaled}
\end{figure}

The main purpose of this proceedings contribution is to achieve a better 
understanding of basic chiral observables in the vicinity of the chiral phase transition temperature 
in (2+1)-flavor QCD and quantify their finite-size
dependence. Although this should eventually been done in the continuum limit of QCD, we stick here to an
analysis of the improved order parameter $M$, introduced in Eqn.~\eqref{eq:M}, on lattices with fixed temporal extent $N_\tau=8$. For our calculations with highly improved staggered quarks and an $\mathcal{O}(a^2)$ Symanzik-improved gauge action, we use the RHMC package embedded within the \textit{SIMULATeQCD} repository \cite{Mazur:2021zgi,HotQCD:2023ghu}. We present here results from our ongoing calculations obtained by using roughly $10K$ statistics or more for each parameter set $(T,m_\ell,V)$ or $(t,H,L)$ in equivalent dimensionless lattice terms. 

In Fig.~\ref{fig:Mscaled}, we show the 
scaled order parameter $M/H^{1/\delta}$
which in the chiral limit, is proportional
to the order parameter scaling function
$f_{G\chi}(z,z_L)$ as discussed in the 
previous section. We show results for 
three temperature values in the vicinity
of the chiral phase transition temperature
on lattices with temporal extent $N_\tau=8$.
This temperature has been estimated previously as $T_c^{N_\tau=8}=143.7(2)$~MeV
\cite{Ding:2024sux}.
We note that the finite-size dependence of $M/H^{1/\delta}$
is close to being linear in the inverse
volume $z_L^3\sim 1/L^3$ (see filled bands in Fig.~\ref{fig:Mscaled}). 
A slight curvature,
however, is visible. This is reminiscent of
the $z_L$-dependence of the $3$-d, $O(2)$
scaling function, which for small $z_L$ has been found to 
be well described by $f_G(0,z_L)\sim z_L^{4}$ in the vicinity of $T_c$ 
\cite{Karsch:2023pga}. We thus fitted
the data for $M/H^{1/\delta}$ obtained for
fixed $H$ as function of $z_{L,b}$, to 
an ansatz,
\begin{eqnarray}
M &=& M_0 \,+\, a_c\, z_{L,b}^c \; ,
\label{fit-ansatz}
\end{eqnarray}
where we (i) either fixed the exponent to $c=3$
or $4$, or (ii) kept $c$ as an additional fit
parameter. The latter ansatz indeed yields
values for $c$ close to 4. 
We also note that the $T$-dependence of the finite volume slope of $M$ is statistically not
significant for the three temperatures shown in
Fig.~\ref{fig:Mscaled}. In fact, fits with fixed $c=4$, as well
as with free exponents $c$,
give amplitudes $a_c$, which agree within errors for the
fits performed at three values of $T$, namely at $T=142.8, 145.1$ and
$147.4$~MeV.

\begin{figure}[htb]
\includegraphics[width=0.45\textwidth]{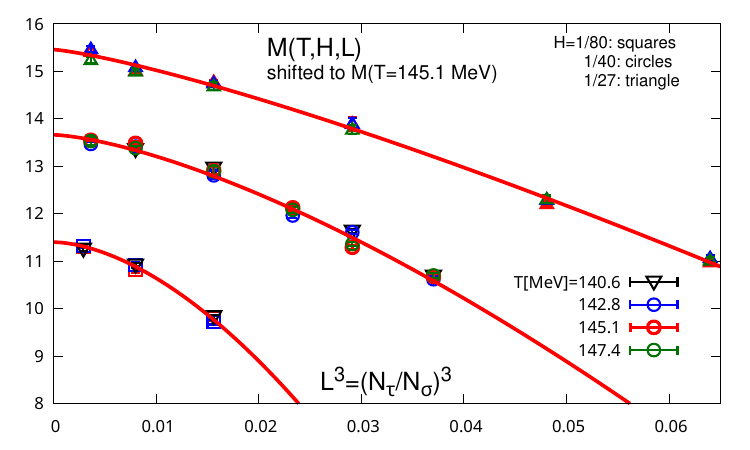}\hspace{1.5cm}
\includegraphics[width=0.45\textwidth]{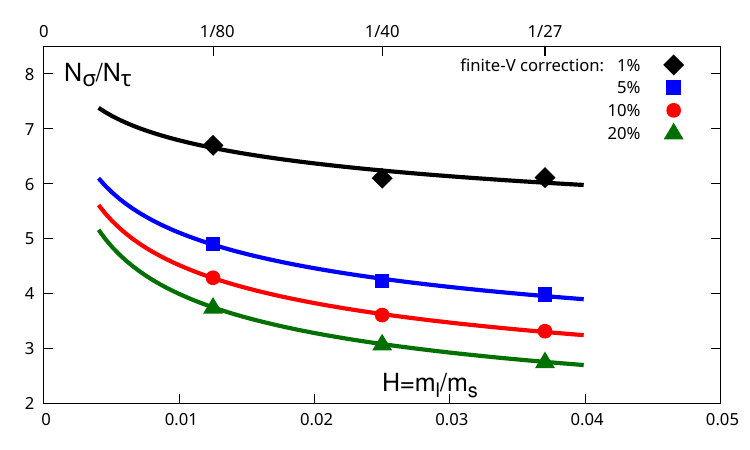}
\caption{{\it Left:} The improved order 
parameter at $T=145.1$~MeV as well as 
at $T=142.8$~MeV and $T=147.4$~MeV. The latter two data sets have been shifted to
the former (see discussion in the text).
{\it Right:} 
Aspect ratio $N_\sigma/N_\tau$ needed to
reduce finite-volume effects in the improved order parameter $M$ below a certain
level, for different light quark masses $m_\ell \equiv m_s H$.}.
\label{fig:Vcorr}
\end{figure}

We therefore performed joint fits to the 
data obtained for all three $T$-values,
using the ansatz given in Eqn.~\eqref{fit-ansatz} with $a_c$ and $c$
being $T$-independent and allowing only 
$M_0$ to be $T$-dependent. Results of 
such a fit are shown in Fig.~\ref{fig:Vcorr}~(left), where we have shifted the data obtained
at $T_1=142.8$~MeV and $T_2=147.4$~MeV, respectively, to $T=145.1$~MeV by using
the difference $s_i=M_0(T_i)-M_0(T)$ as shift
parameters. Results obtained for the fits
shown in Fig.~\ref{fig:Vcorr}~(left) are
given in Table~\ref{tab:fit}.

We may use these fits to estimate the aspect ratio $N_\sigma/N_\tau$\,, that is needed in
lattice QCD calculations at fixed $H$ to
reduce finite-size effects in calculations of the improved order parameter below a certain percentage level. Such estimates
are shown in Fig.~\ref{fig:Vcorr}~(right).
This figure shows that in order to reduce finite-size effects on $M$ below the one percent 
level, aspect ratios larger than $6$ are needed for
physical light-to-strange quark mass ratios, $H=1/27 \sim 0.037$ while an aspect ratio of about $7.5$ is needed
at $H=1/80=0.0125$, which corresponds to a Goldstone pion
mass of about $80$~MeV. We also note that the 
fits performed here are valid for \,$z_{L,b}\,\,\lsim \,\,1.5$.
For physical quark mass values or $H=1/27$, this corresponds to an aspect ratio
$N_\sigma/N_\tau\simeq 2.5$. For larger values of $z_{L,b}$, 
higher order corrections in $z_{L,b}$ need to be 
taken into account for an accurate description of
the finite-size dependence of the order parameter. 

\begin{table}[htb]
\begin{center}
\begin{tabular}{|c|c|c|c|c|c|}
\hline
	$H$ & $M_0(T)$ & $a_c$ & c &$s_1$&$s_2$ \\
\hline
	1/27 & 15.46(7)& -0.95(8) & 3.77(18)& 1.21(6) & -1.32(7) \\
	1/40 & 13.66(5)& -0.56(8) & 4.39(37)& 1.42(6) & -1.50(8) \\
	1/80 & 11.40(4)& -0.23(4) & 5.16(43)& 1.68(3) & -1.96(3)\\
\hline
\end{tabular}
\end{center}
	\caption{Results for
	fits of $M/H^{1/\delta}$ using the ansatz given in Eqn.~\eqref{fit-ansatz} with
	$M_0(T)$ for $T=145.1$~MeV and $M_0(142.8 {\rm MeV}) =M_0(T)+s_1$, $M_0(147.4 {\rm MeV}) =M_0(T)+s_2$. The data for $T_1=142.8$~MeV and $T_2=147.4$~MeV have been shifted by $s_1$ and $s_2$, respectively. 
}
\label{tab:fit}
\end{table}

\begin{figure}[htb]
\begin{center}
\includegraphics[width=0.7\textwidth]{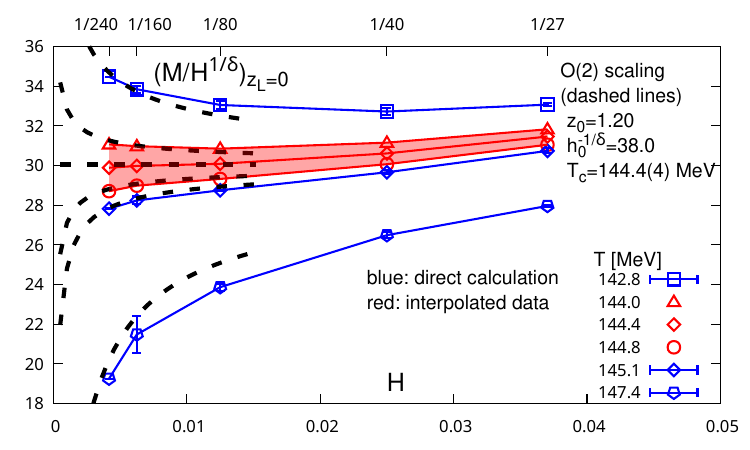}
\caption{The infinite volume extrapolated, scaled order parameter versus 
$H$ for three values of the temperature 
(blue) and interpolated values (red) used
to estimate the chiral phase transition temperature on lattices with temporal extent $N_\tau=8$. 
Dashed lines show the infinite volume, asymptotic scaling results using the 
non-universal parameters given in the figure.
}
\label{fig:Tcestimate}
\end{center}
\end{figure}

Finally, we want to use the infinite volume extrapolated results for $M/H^{1/\delta}$ to obtain a
new estimate for the chiral phase transition 
temperature on lattices of temporal extent $N_\tau=8$. In Fig.~\ref{fig:Tcestimate}, we show
the scaled order parameter $M/H^{1/\delta}$ versus $H$ for the three
temperature values examined so far. In addition 
to the infinite volume extrapolation of the data 
for $H\ge 1/80$ shown in Fig.~\ref{fig:Mscaled},
we also added some preliminary results obtained for 
$H=1/160$ and $H=1/240$. The tendency for $M/H^{1/\delta}$ to diverge in chiral limit at $T=142.8$~MeV and vanish at $T=147.4$~MeV is clearly observed, following theoretical expectations based on the asymptotic expansion of the scaling functions. At $T=145.1$~MeV, the data continues to decrease for reducing value of 
$H$ in chiral limit, suggesting that this temperature value also
is above $T_c^{N_\tau=8}$. 

Also shown in Fig.~\ref{fig:Tcestimate} are results
for $M/H^{1/\delta}$ obtained from a linear interpolation of data at $T=142.8$~MeV and $T=145.1$~MeV, respectively. Demanding that the
interpolated data at $H=1/160$ and $1/240$ agree
within errors, yields an estimate for $T_c^{N_\tau=8}$. As our data generation and analysis
is not yet complete for the two lowest values of $H$,
the new estimate of the chiral transition temperature based on
a well-controlled infinite volume extrapolation, also is still preliminary. At present, we find $T_c^{N_\tau=8}=144(4)$~MeV, which is in reasonable 
agreement with earlier estimates \cite{Ding:2024sux}.

Using this estimate for $T_c$ and the value 
$h_0^{-1/\delta}$ obtained as,
\begin{equation}
    h_0^{-1/\delta} =\lim_{H\rightarrow 0}\left[ \frac{M(144~{\rm MeV},H)}{H^{1/\delta} \left(1-\frac{1}{\delta}\right)}\right] \simeq 38 \; , 
\end{equation}
we can determine the only remaining non-universal 
parameter $z_0$ from a fit to the three values for
$M/H^{1/\delta}$ at $H=1/240$. We obtain this as $z_0\simeq 1.2$. The resulting asymptotic scaling curves are shown as dashed lines in Fig.~\ref{fig:Tcestimate}. As can be seen, the asymptotic scaling curve $f_{G\chi}(z)$ gives a 
fairly good approximation to the scaled order parameter data for $H\le 1/160$ and in an even larger
range, $H\le 1/80$, at as well as close to $T_c$.

\section{Summary and Outlook}
\label{sec:Summary}

We presented first results from an ongoing study
of finite-size effects in calculations of an improved
order parameter for chiral symmetry breaking in $(2+1)$-flavor QCD, in which we attempt to quantify the finite-size effects as function of lattice volume and the light quark masses for a given temperature close to the chiral phase transition. Our calculations on lattices
with fixed lattice cut-off corresponding to a temporal extent of $N_\tau=8$ show that precision studies of the order parameter require aspect ratios for the lattice size in units of temperature, $N_\sigma/N_\tau\ge 6$ at physical values of the 
quark masses, which increases to about $8$ for the currently used
smallest quark mass ratio $H=m_\ell/m_s=1/240$. This is currently a work in progress.

We also show that the improved order parameter $M$
receives corrections to the leading order universal
critical behavior from corrections-to-scaling and/or
regular terms for $m_\ell/m_s>1/160$. These corrections stay below the 10\% level at the physical value of the light-to-strange quark mass ratio $H=1/27$.

A more detailed, quantitative analysis of corrections
to leading order universal scaling does require
further calculations at small values of  $m_\ell/m_s$\,,
which are ongoing.

\acknowledgments

This work is supported by the Deutsche Forschungsgemeinschaft, German Research Foundation, Proj. No. 315477589-TRR 211. We sincerely acknowledge the computing time made available on the following : (i) clusters Noctua2 and Otus at the NHR Center Paderborn Center for Parallel Computing (PC2), (ii) supercomputer JUPITER at Jülich Supercomputing Centre (JSC) by the Gauss Centre for Supercomputing e.V., (iii) EuroHPC supercomputer LEONARDO hosted by CINECA (Italy) through the EuroHPC Joint Undertaking and (iv) Bielefeld HPC-GPU clusters in Bielefeld University. We also sincerely thank Olaf Kaczmarek, Christian Schmidt, Mugdha Sarkar and Sipaz Sharma for useful discussions and their contributions to this ongoing research project.

\bibliographystyle{JHEP}
\bibliography{bibliography}

\end{document}